\documentclass{vgtc}                          

\ifpdf
  \pdfoutput=1\relax                   
  \usepackage{graphicx}                
  \DeclareGraphicsExtensions{.pdf,.png,.jpg,.jpeg} 
\else
  \usepackage{graphicx}                
  \DeclareGraphicsExtensions{.eps}     
\fi%

\graphicspath{{figures/}{pictures/}{images/}{./}} 

\usepackage{microtype}                 
\PassOptionsToPackage{warn}{textcomp}  
\usepackage{textcomp}                  
\usepackage{mathptmx}                  
\usepackage{times}                     
\usepackage{cite}                      
\usepackage{tabu}                      
\usepackage{booktabs}                  
\usepackage{listings}
\usepackage{glslListings}
\usepackage{hyperref}
\usepackage[usenames, svgnames]{xcolor}
\usepackage{url}

\newcommand{\insitu}{\textit{in situ}~}
\newcommand{\scenery}{\textit{scenery}~}
\usepackage{contour}
\usepackage[normalem]{ulem}

\contourlength{0.8pt}

\newcommand{\myuline}[1]{%
  \uline{\phantom{#1}}%
  \llap{\contour{white}{#1}}%
}

\onlineid{0}

\vgtccategory{Research}

\vgtcinsertpkg

\title{A Proposed Framework for Interactive Virtual Reality \textit{In Situ} Visualization of Parallel Numerical Simulations}

\author{Aryaman Gupta\thanks{e-mail: argupta@mpi-cbg.de}%
\hspace{4em} Ulrik G\"{u}nther%
\hspace{4em} Pietro Incardona%
\hspace{4em} Ata Deniz Aydin\\%
\parbox{5in}{\scriptsize \centering Technische Universit\"at Dresden \& Center for Systems Biology Dresden (CSBD) \& MPI-CBG, Dresden} %
\\[1em]
\and Raimund Dachselt\\ %
     \parbox{1.8in}{\scriptsize \centering Technische Universit\"at Dresden\\ Interactive Media Lab Dresden \\ Cluster of Excellence Physics of Life}%
\and Stefan Gumhold\\ %
     \parbox{1.8in}{\scriptsize \centering Technische Universit\"at Dresden \\ Cluster of Excellence Physics of Life}%
\and Ivo F.~Sbalzarini\thanks{e-mail: sbalzarini@mpi-cbg.de}\\ %
     \parbox{1.8in}{\scriptsize \centering Technische Universit\"at Dresden \\ CSBD \& MPI-CBG, Dresden \\ Cluster of Excellence Physics of Life}
     \vspace{-5mm}
     }

\abstract{As computer simulations progress to increasingly complex, non-linear, and three-dimensional systems and phenomena, intuitive and immediate visualization of their results is becoming crucial. While Virtual Reality (VR) and Natural User Interfaces (NUIs) have been shown to improve understanding of complex 3D data, their application to live \insitu visualization and computational steering is hampered by performance requirements. Here, we present the design of a software framework for interactive VR \insitu visualization of parallel numerical simulations, as well as a working prototype implementation. Our design is targeted towards meeting the performance requirements for VR, and our work is packaged in a framework that allows for easy instrumentation of simulations. Our preliminary results inform about the technical feasibility of the architecture, as well as the challenges that remain. 
} 

\CCScatlist{
  \CCScatTwelve{Computing methodologies}{Massively parallel and high-performance simulations}{}{};
  \CCScatTwelve{Human-centered computing}{Virtual reality}{}{};
}

\begin{document}


\firstsection{Introduction}

\maketitle

The growing disparity between computational throughput and I/O bandwidth to disk increasingly prohibits the classical approach of writing simulation output to disk and analyzing it in  post-processing steps. {\textit{In situ}} techniques, which involve the visualization or analysis of a simulation as it runs, have thus received much research interest.

In addition to helping bypass the file I/O bottleneck, \insitu visualization also enables computational steering. Images rendered \insitu may be streamed to a remote client, where a user can visualize the simulation live and also interact with it. This has, e.g., been applied in computational fluid dynamics to interactively vary simulation parameters~\cite{sanderson2018coupling}, which can help gain an intuition of the simulated physics. 

Forming an intuition of the simulated dynamics and its parameter dependence benefits from Virtual Reality (VR) visualization, which has been shown to improve perception of space and geometry~\cite{slater2016enhancing}. We therefore believe that \insitu VR visualization has the potential to aid scientific discovery by providing better understanding of simulation data, and an environment for intuitive interaction with it. 

A VR environment lends itself to the use of Natural User Interfaces (NUI) for intuitive interaction. NUIs are based on how humans interact with their physical environment and include the use of body pose, hand gestures, hand-held tools, and eye gaze. It has been shown that scientists better understand data when they are able to interact with it via intuitive interfaces~\cite{brooks1990project}. 

Using VR for \insitu visualization of simulations, however, is complicated by the high frame rates and low latency required. Simulation data can be large, often tens or hundreds of gigabytes per simulation time step, which typically occurs once per second. \textit{In situ} visualization in VR is therefore difficult to achieve.

Here, we propose a framework for \insitu VR visualization and NUI steering of simulations that optimizes rendering frame rate and latency. We describe the architecture and present preliminary results, which leads us to identifying challenges that need to be overcome in future work.

\section{Rendering and Simulation Libraries Used}
The rendering system used in our framework is developed using \scenery \cite{gunther2019scenery}, an open-source VR library for both volumetric and geometric data. In simulations, volumetric data typically stem from discretized continuous fields and geometric data from surface meshes or postprocessing results, such as isosurfaces or streamlines. \scenery supports rendering to both commodity VR headsets and distributed setups, such as VR CAVEs and PowerWalls. It uses the high-performance Vulkan API for rendering, and therefore has the potential to harness the power of modern GPUs better than traditional OpenGL-based solutions, as still predominantly used by \insitu solutions like ParaView Catalyst \cite{ayachit2015paraview} and VisIt Libsim \cite{kuhlen2011parallel}. 

The simulation engine used in our framework is OpenFPM \cite{incardona2019openfpm}, an open-source library that enables  rapid development of scalable and efficient particle- and mesh-based numerical simulations. OpenFPM largely automates parallelization on high-performance computers, including distributed-memory systems, many-core systems, and GPGPUs. It has been used to implement simulations ranging from Smoothed-Particle Hydrodynamics to Molecular Dynamics to Discrete Element Methods on thousands of CPU cores~\cite{incardona2019openfpm}.

Integrating our work into OpenFPM helps us reach a wide audience of computational scientists who benefit from interactive \insitu visualization and steering, but may lack the time or expertise to write scalable solutions from scratch. We  implement \insitu functionality into OpenFPM in a manner that allows users to enable it in a fraction of the time and effort it normally takes to instrument a simulation with an \insitu solution.

\section{In Situ Architecture}\label{sec:framework}

We design our \insitu architecture targeting a rendering frame rate of at least 60 Hz, rendering latency on viewpoint changes of at most 20~ms, and low interaction latency for steering commands.

\begin{figure}[tb]
 \centering 
 \includegraphics[width=\columnwidth]{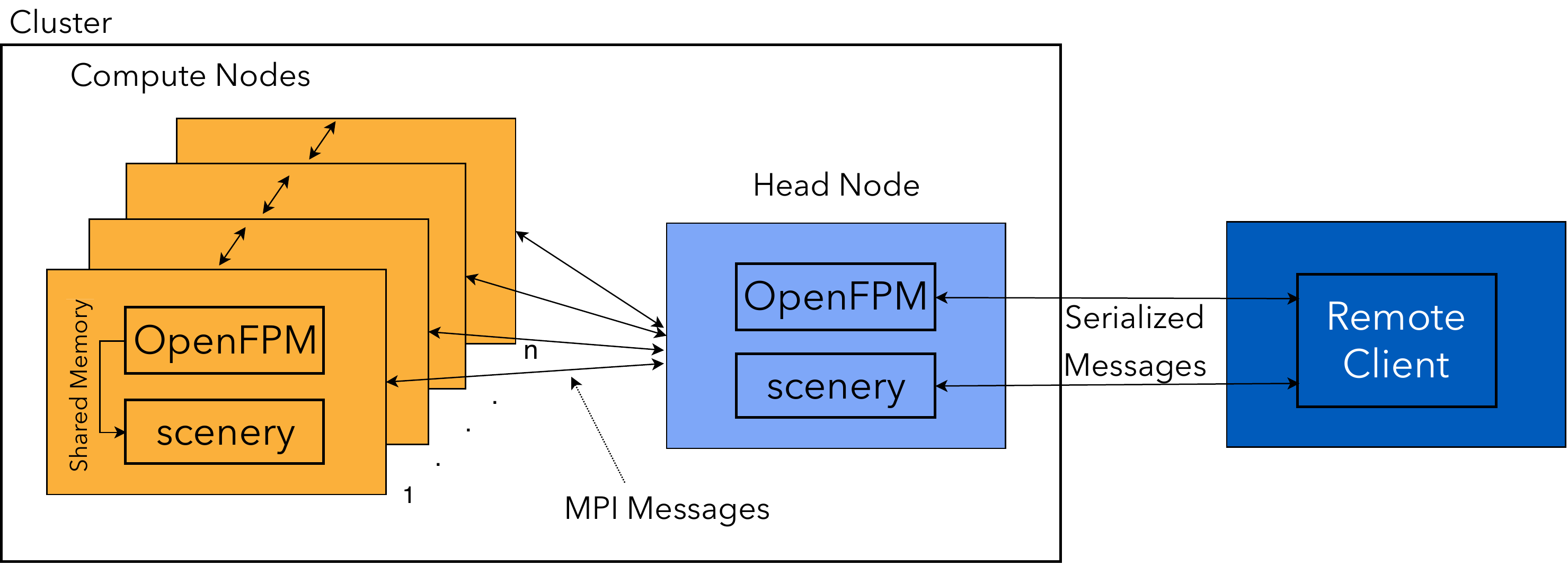}
 \caption{Schematic of our framework architecture.}
 \label{fig:sample}
\end{figure}

Key to achieving these goals is to minimize data movement. We thus choose a tightly-coupled structure with simulation and rendering sharing resources, running on the same nodes within a cluster. Figure~1 shows a schematic. Each compute node runs the distributed OpenFPM environment for the simulation, along with a \textit{scenery}-based application that performs local rendering. All simulation data generated on a node are rendered on the node itself. The simulation data are shared between OpenFPM and \scenery via a common pointer to shared memory. This enables zero-copy rendering.

Several \insitu solutions, including ParaView Catalyst \cite{ayachit2015paraview} and VisIt Libsim \cite{kuhlen2011parallel}, adopt a synchronous execution strategy, where simulation and visualization run sequentially. We choose an asynchronous strategy, aiming to reduce interaction and rendering latency. In our asynchronous strategy, neither simulation nor visualization need to wait for the other, which means that steering commands can be incorporated at the earliest, and rendering can begin while the simulation is in progress. Visualization artifacts that may result from this lack of synchronization are expected to be small, since numerical stability requires the simulation to only change slightly between time steps. For our application, we emphasize low latency over these minor visual artifacts.

The rendering results produced on the compute nodes are then composited using a divide-and-conquer strategy similar to the binary swap algorithm \cite{ma1994parallel}. The \scenery process on the Head Node is responsible for streaming the composited result to a remote client, as well as for receiving requests for changes in visualization parameters. The remote client also communicates steering commands to the OpenFPM process on the Head Node, which conveys them to all OpenFPM processes via MPI messages.

In contrast to other approaches, we intend to render an explorable representation, such as a Volumetric Depth Image (VDI)~\cite{frey2013explorable}, instead of a simple image. Such representations of volumetric data can, once generated, be reprojected to a new viewpoint without redoing the ray casting. We intend to render such a representation on the cluster and stream it to the remote client. Small viewpoint changes can then be handled locally by the remote client without involving network latency. This is crucial for meeting the latency requirements of VR, where viewpoint changes occur frequently.

\section{Preliminary Results}

We have implemented a prototype of our framework, performing particle-as-sphere rendering of a distributed OpenFPM simulation. The simulation runs parallel on multiple CPUs, while visualization is done on the GPU. The shared-memory communication between the processes on a single node is extended by a protocol to handle dynamic memory requests and reallocations. The \scenery application reads the simulation data from shared memory and understands the data structures used by OpenFPM. Our implementation is open source. The \scenery application can be found at \href{https://github.com/scenerygraphics/scenery-insitu}{\color{MidnightBlue}{\myuline{github.com/scenerygraphics/scenery-insitu}}} and the OpenFPM project at \href{https://git.mpi-cbg.de/openfpm/openfpm_pdata}{\color{MidnightBlue}{\myuline{git.mpi-cbg.de/openfpm/openfpm\_pdata}}} (in the \mbox{``insitu\_visualization''} branch).

Figure~2 shows our prototype applied to a Molecular Dynamics simulation. The simulation code written using OpenFPM is only 161 lines of C++, and the inclusion of \insitu visualization requires changes in only two lines.

\begin{figure}[tb]
    \centering
    \includegraphics[width=\columnwidth]{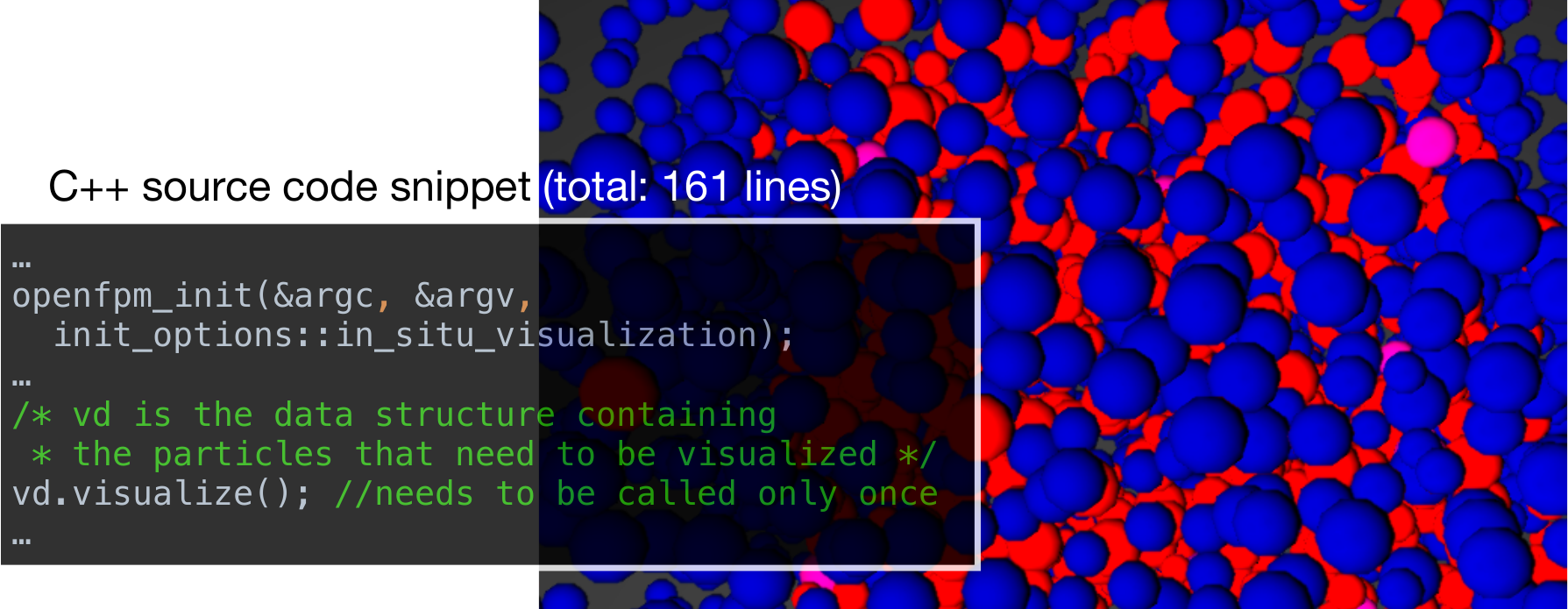}
    \caption{An OpenFPM-based distributed Molecular Dynamics simulation visualized \insitu using our prototype, and the lines of code that needed to be changed to enable \insitu visualization. Particles are colored by velocity magnitude.}
    \label{fig:mdFigure}
\end{figure}


\section{Future Work and Conclusions}
We have presented a framework architecture for interactive VR \insitu visualization of parallel numerical simulations. We used \scenery and OpenFPM to base the rendering and simulation parts of the framework on. We have implemented a first prototype of the system, demonstrating the feasibility of achieving our design goals. In our prototype, enabling live \insitu visualization only required minimal changes in the existing simulation code.

Our prototype does not yet meet the targets we set for immersive \insitu visualization in Section \ref{sec:framework}, but it enables us to identify a number of challenges that should be addressed in order to meet them: 
First, volumetric rendering results need to be in an explorable representation, such as a VDI, as explained in Section \ref{sec:framework}. Second, we need a solution to efficiently generate and composit such a representation on a cluster. Third, support needs to be added for simulations running on GPGPUs. In addition, we are going to extend our framework to NUI-based steering by instrumenting OpenFPM to receive steering commands on the fly and by expanding  \textit{scenery}'s support for NUI interaction.

Taken together, we are confident that many applications in scientific computing will benefit from a more immersive and more explorative mode of interaction.


\bibliographystyle{abbrv-doi}

\bibliography{template}
\end{document}